\documentstyle[preprint,revtex]{aps}
\begin{document}
\begin{title}
A solvable spin glass of quantum rotors.
\end{title}
\author{J. Ye, S. Sachdev and N. Read}
\begin{instit}
Departments of Physics and Applied Physics, P.O. Box 2157\\
Yale University, New Haven, CT 06520
\end{instit}
\begin{abstract}
We examine a model of $M$-component quantum rotors coupled by
Gaussian-distributed random, infinite-range exchange interactions.
A complete solution is obtained at $M=\infty$ in the
spin-glass and quantum-disordered phases.
The quantum phase transition separating them is found to possess logarithmic
violations of scaling, with no further modifications to the leading critical
behavior at any order in $1/M$; this suggests that the critical
properties of the
transverse-field Ising model (believed to be identical to the
$M\rightarrow 1$ limit)
are the same as those of the $M=\infty$ quantum rotors.
\end{abstract}
\narrowtext

Extensive attention has been lavished in the last decade on the problem of
classical spin glasses and their finite temperature ($T$) phase transition to a
spin glass ordered phase~\cite{hertz}. In contrast, there has been relatively
little work on quantum spin glasses~\cite{bm1,gold1,gold2},
especially on their $T=0$ quantum phase
transition to a quantum disordered (or `spin-fluid') state.
In particular, there is no model for which the critical properties of this
quantum phase transition have been obtained. On the experimental side, there
has
been a renewed interest in a number of spin systems which are in the
vicinity of a $T=0$ phase transition from a spin-glass to a spin-fluid
state~\cite{rosen,birg,gabe,qcrit,kagome}: these include the dipolar,
transverse-field Ising magnet $Li Ho_x Y_{1-x} F_4$~\cite{rosen},
the lightly-doped cuprates~\cite{birg,gabe,qcrit}, and various layered
transition-metal/rare-earth oxides~\cite{kagome}.

In this paper we examine a quantum spin glass which allows us to examine
more carefully the nature of the quantum spin-glass to spin-fluid phase
transition and determine the spectrum of excitations in the
spin-fluid phase. We consider $M$-component quantum rotors with
Gaussian-distributed random,
infinite-range exchange interactions. A complete solution of this model will
be obtained at $M=\infty$ in both the spin-glass and spin-fluid phases
and at the critical point separating them. We also examine the nature
of the $1/M$ corrections at $T=0$ in the spin-fluid phase and at the critical
point: we find that the form of the leading critical behavior and
the low-frequency spectral weight remains unmodified to all orders in $1/M$
from
the $M=\infty$ result. Thus the results of this paper
could have been derived without any reference
to the $1/M$ expansion, by simply resumming Feynman graphs which are dominant
at low frequency - these graphs happen to be identical to those selected
by the $M=\infty$ theory.

The quantum rotors should not be confused with true quantum Heisenberg spins
present in any isotropic antiferromagnet; 
the different components of the
rotor variables all commute with each other, unlike the quantum spins.
As a consequence, the path-integral written in the rotor variables has an
action which contains no Berry phases and is purely real.
The properties of random quantum spin models are quite different from those
of the quantum rotors considered here, and
will be discussed elsewhere~\cite{heisen}.
Apart from its
theoretical simplicity, the main utility of the rotor model is that the
$M=1$ limit of the path integral is expected to be in the same universality
class as the Ising model in a transverse field.
The absence of any $1/M$ corrections noted above, suggests that
the critical-behavior of the infinite-range,
transverse-field Ising model is identical to that of the $M=\infty$ limit
solved in this paper. This is also consistent with a recent analysis
of this Ising model by Huse and Miller~\cite{huse}: their results for
the critical point are essentially identical to those obtained below in the
$M=\infty$ model.

We will study the following ensemble of Hamiltonians
\begin{equation}
H = \frac{g}{2M} \sum_{i} \hat{L}^2 +
\frac{M}{\sqrt{N}} \sum_{i<j }J_{ij} \hat{n}_{i} \cdot \hat{n}_{j}
{}~~~~~~\hat{n}_{i}^2 = 1
\end{equation}
where $i,j$ extend over $N$ sites, $n_{i\mu}$ are the $M$ components of a
unit-length
rotor $\hat{n}_i$ on site $i$, the $L_{i\mu\nu}$ ($\mu < \nu$, $\mu,\nu = 1
\ldots M$) are the
$M(M-1)/2$ components of the angular-momentum
generator $\hat{L}_i$ in rotor space, and the $J_{ij}$ are
mutually uncorrelated exchange constants selected with probability
$P(J_{ij}) \sim \exp(-J_{ij}^2 /(2 J^2))$. The $n_{i\mu}$ are
mutually commuting variables and the quantum dynamics is
defined by the commutation relations:
\begin{equation}
[ L_{i\mu\nu} , n_{j\sigma} ] = i\delta_{ij}(\delta_{\mu\sigma}n_{j\nu}
-\delta_{\nu\sigma} n_{j\mu} )~~~~~
\end{equation}
The $L_{i\mu\nu}$ satisfy the commutation relations of angular momenta in
$M$ dimensions.
As $g\rightarrow 0$, the model reduces to the {\em classical}, infinite-range,
$M$-component, Heisenberg spin glass which was analyzed earlier by de Almeida
{\em et.al.\/}~\cite{almeida}.

The formulation of the $N\rightarrow \infty$ limit of $H$ can be obtained
by a straightforward generalization of the analyses in
Refs~\cite{bm1,gold1,gold2}. We use the path-integral formulation of the
partition function, introduce $n$ replicas, and average over the ensemble of
the $J_{ij}$. The $N\rightarrow\infty$ limit yields a saddle-point which
describes the quantum mechanics of $n$ replicas of a single rotor.
Assuming the saddle-point is $O(M)$ invariant (this is true in both
the spin-fluid and spin-glass phases) we obtain the single-site
path-integral
\widetext
\begin{equation}
Z_0 = \int {\cal D} \hat{n}^{a} ( \tau ) \delta\left(
\hat{n}^{a2} ( \tau ) - 1\right)
\exp \left(
- \frac{M}{2g} \int_0^{\beta} d\tau (\partial_{\tau} \hat{n}^{a})^2
+ \frac{MJ^2}{2} \int_0^{\beta} d \tau d \tau^{\prime}
Q^{ab} ( \tau - \tau^{\prime}) \hat{n}^{a} ( \tau ) \cdot
\hat{n}^{b} ( \tau^{\prime} ) \right)
\label{z0}
\end{equation}
\narrowtext
and the self-consistency condition
\begin{equation}
Q^{ab} ( \tau - \tau^{\prime} ) =  \langle \hat{n}^{a} ( \tau ) \cdot
\hat{n}^{b}
( \tau^{\prime} ) \rangle_{Z_0}
\label{qab}
\end{equation}
Here $a,b = 1 \ldots n$ are replica indices, $\tau$, $\tau^{\prime}$ are
Matsubara times, and $\beta = 1/T$.
The Edwards-Anderson order parameter~\cite{hertz} for the spin-glass phase
is
\begin{equation}
q_{EA} = Q^{aa} ( \tau
\rightarrow \infty ).
\end{equation}
Moreover,
$Q^{ab}$, $a \neq b$, is $\tau$-independent and
non-zero only in the spin-glass
phase~\cite{gold2}.

An exact evaluation of $Z_0$ is clearly not possible. We present below the
results of a systematic $1/M$ expansion on $Z_0$.

\underline{$M=\infty$ theory}: Imposing the constraint by a Lagrange-multiplier
$\lambda$, the $M=\infty$ limit of Eqns (\ref{z0},\ref{qab}) reduces
to the constraint $Q^{aa} ( \tau = 0 ) = 1$ and
\begin{equation}
Q ( i \omega_n ) = g \left(\omega_n^2 + \lambda - g J^2 Q ( i \omega_n )
\right)^{-1}
\label{selfcon}
\end{equation}
where $Q(i\omega_n)$ is the Fourier transform of $Q(\tau)$ at the Matsubara
frequencies, and the r.h.s is a matrix inverse in replica space.
\newline
{\em 1. Paramagnetic phase:\/} For large $g$, or large $T$, we expect a
paramagnetic phase (the quantum-disordered phase is the $T=0$ paramagnetic
state)
in which case $Q^{ab}$ will be replica diagonal~\cite{gold1,gold2}.
A closed-form solution can be obtained from (\ref{selfcon})
for the spectral weight
$\chi^{\prime\prime} ( \omega ) = \mbox{Im} ( Q^{aa} ( \omega + i 0^{+} ) )$:
\begin{equation}
\chi^{\prime\prime} ( \omega ) = \mbox{sgn} ( \omega ) \frac{
\left[ ( \omega^2 - \lambda + 2 J g ) ( \lambda + 2 J g - \omega ^2 )
\right]^{1/2}}{2J^2 g}
\label{chires}
\end{equation}
for $\lambda-2Jg < \omega^2 < \lambda + 2 Jg$
and $\chi^{\prime\prime} = 0 $ otherwise. It is clear that a physically
sensible solution requires $\lambda \geq 2 J g$ where $\lambda$ is determined
by the constraint equation $\hat{n}^{a2} = 1$, or
\begin{equation}
\int_0^{\infty} \frac{d \omega }{\pi} \chi^{\prime\prime} ( \omega )
\mbox{coth} ( \beta \omega /2) = 1
\label{intchi}
\end{equation}
It is evident from (\ref{chires}) that the $M=\infty$ paramagnet has a
gap of $(\lambda - 2 J g )^{1/2}$ towards spin-wave excitations. We expect
$1/M$ fluctuations to fill in this gap at any finite $T$; the gap in
the $T=0$ spin-fluid phase is however robust towards such corrections.
The paramagnetic-spin glass phase boundary is determined by setting
$\lambda = 2 J g$ and solving (\ref{intchi}) for a line in the $g-T$ plane: the
results of this calculation are shown in Fig~\ref{phase}. The quantum
transition near $T=0$ occurs at $g = 9 \pi^2 J/16 - 3T^2 /J + \cdots$,
and the classical transition near $g=0$
occurs at $T = J - g/12 + \cdots$;
this latter result agrees with that of Ref.~\cite{almeida}.
\newline
{\em 2. Spin-glass phase:\/} We now expect {\em only\/} $Q^{ab} ( i\omega_n =
0 )$
to acquire off-diagonal components~\cite{gold1,gold2}; the finite-frequency
$Q ( i\omega_n ) $
remains diagonal. We therefore parametrize
\begin{equation}
Q^{aa} ( i\omega_n ) = Q^{aa}_{reg} ( i\omega_n ) +  \beta q_{EA}
\delta_{\omega_n, 0}
\end{equation}
where $Q^{aa}_{reg} ( i\omega_n )$ can be obtained immediately from the
solution
of (\ref{selfcon}) and continues to have spectral weight
$\chi_{reg}^{\prime\prime} ( \omega )$ which obeys (\ref{chires}) with a value
of $\lambda$ to be determined below.
We parametrize the off-diagonal components of $Q^{ab} ( i\omega_n = 0 )$
by an arbitrary hierarchical matrix~\cite{parisi} specified by a
monotonic function $\beta q(x)$ on the interval $0\leq x \leq 1$.
Using the expressions for the inverse of an hierarchical matrix
in Ref.~\cite{mezard}, the self-consistency equation (\ref{selfcon})
can be transformed into
two integral equations for $q(x)$ and $q_{EA}$. Simple algebraic manipulations
then yield the satisfactory~\cite{hertz} result
\begin{equation}
q_{EA} = q(1)
\end{equation}
Repeated differentiation of the integral equations showed that
$dq/dx = 0$; $q(x)$ can therefore only be a piecewise constant function.
We chose $q(x) = q_1$ for $0<x<u$ and $q(x) = q_{EA}$ for $u < x < 1$,
whence the integral equations specified $q_1 = 0 $ and $q_{EA}$; $u$ was
however
left undetermined~\cite{mezard}. It was then necessary to evaluate the free
energy and demand stationarity with respect to $u$. The final result was
quite simple: we found $u=0$ implying that $q(x) = q_{EA}$ for all $x$ and that
the replica-symmetric solution is optimal. This agrees with the classical
limit at $g=0$ which was found in Ref.~\cite{almeida} to possess a stable
replica symmetric solution at $M=\infty$; we also undertook a stability
analysis, similar to that in Ref.~\cite{almeida},
for the quantum-rotor model and found only non-negative eigenvalues
in the fluctuations about the replica-symmetric state.
Our final results for the spin-glass phase were: $\lambda = 2 Jg$ with
$\chi^{\prime\prime}_{reg} ( \omega )$ given by (\ref{chires}) being
gapless over the entire phase,
$Q^{ab} ( i\omega_n = 0 ) = \beta q_{EA}$ for $a\neq b$ and
\begin{equation}
q_{EA} = 1 - \int_0^{\infty}
\frac{d \omega }{\pi} \chi^{\prime\prime}_{reg} ( \omega )
\mbox{coth} ( \beta \omega /2)
\label{qeares}
\end{equation}
{\em 3. Quantum critical region:\/} We now examine the region near the
quantum phase transition at $g=g_c \equiv 9 \pi^2 J/16$, $T=0$.
Scaling (see {\em e.g\/} Ref.~\cite{qcrit}) predicts
that the spin-glass paramagnetic boundary obeys $T \sim |\delta g|^{z\nu}$
(here $\delta g \equiv g - g_c$). From the equation for the phase-boundary
at small $T$ above, we deduce $z \nu = 1/2$.
The order-parameter $q_{EA}$ must vanish as $q_{EA} \sim |\delta g|^{\beta}$;
from (\ref{qeares}) this yields $\beta = 1$. Further the $T=0$ spin-gap,
$\Delta$, in the quantum-disordered phase
should vanish as $\Delta \sim (\delta g)^{z\nu}$. Using
$\Delta = (\lambda - 2 Jg)^{1/2}$ and (\ref{intchi}), we find however
that $\Delta \sim (\delta g/\mbox{log}(1/\delta g))^{1/2}$. Thus there is a
surprising logarithmic violation of naive scaling - the log divergence is
a consequence of the square-root threshold in the spectral weight
(\ref{chires}).
For $\omega$ and $\delta g$ small, but $\omega/\delta g$ arbitrary,
the entire $T=0$, local dynamic susceptibility obeys a scaling form:
\begin{equation}
\chi^{\prime\prime} ( \omega , T=0 ) = c_1 \mbox{sgn}(\omega) | \omega |^{\mu}
\Phi_g \left( \frac{\omega}{\Delta_g} \right)
\end{equation}
where the frequency scale $\Delta_g$ obeys
$\Delta_g = c_2 (\delta g)^{zv}/\log^{1/2}(1/\delta g)$ for small $\delta g$,
the exponent $\mu = -1 +
\beta /(z \nu ) = 1$~\cite{qcrit},
$c_1 , c_2$ are non-universal constants, and $\Phi_g$ is a universal function
given by
\begin{equation}
\Phi_g ( x ) = \left\{ \begin{array}{cc} (1-1/x^2 )^{1/2} & \mbox{for $|x| >
1$}
\\ 0 & \mbox{otherwise} \end{array} \right.
\end{equation}
We will argue below that the results for $z\nu$, $\beta$, $\mu$ and
$\Phi_g$ are in fact exact to all orders in $1/M$; only the non-universal
constants $c_1 , c_2$ get modified by higher order corrections.
A related analysis can be performed at the critical coupling $g=g_c$ but at
finite temperature~\cite{qcrit}. For $\omega$ and $T$ small, but with
$\omega /T$ arbitrary, the local dynamic susceptibility now obeys the
scaling form:
\begin{equation}
\chi^{\prime\prime} ( \omega , g=g_c ) = c_1 \mbox{sgn}(\omega) | \omega
|^{\mu}
\Phi_T \left(  \frac{\hbar \omega }{\Delta_T}
 \right)
\end{equation}
where
the universal
function $\Phi_T$ is
\begin{equation}
\Phi_T ( x ) = \left\{ \begin{array}{cc} (1-4\pi^2/(3x^2) )^{1/2}
& \mbox{for $|x| > 2\pi/\sqrt{3}$}
\\ 0 & \mbox{otherwise} \end{array} \right.
\end{equation}
and the frequency scale $\Delta_T = k_B T / \log^{1/2} ( 1/T )$ at low $T$,
with {\em no non-universal prefactor.} Note again the presence of logarithmic
violations of naive scaling; the frequency-scale for the dynamic susceptibility
is however still set completely by the absolute temperature to leading-log
accuracy.
The presence of a gap in $\Phi_T$ is clearly an artifact of the
large $M$ limit~\cite{qcrit}, as the $T=0$ state is gapless at $g=g_c$;
we expect $1/M$ corrections to modify $\Phi_T$ by filling in the gap.

\underline{$1/M$ expansion:} We now examine corrections to the above mean field
theory at $T=0$ in the quantum-disordered phase and at the
quantum-critical point, $g=g_c$.
We will not examine such corrections in the spin-glass phase where the
structure is considerably more complicated due to the expected
appearance of replica symmetry breaking.
Our main result will be that neither the critical exponents nor the
form of the low frequency spectral weights are modified by the $1/M$
corrections.
We begin by absorbing all higher-order corrections
into a self-energy, $\Sigma$, in the $\hat{n}$ propagator,
which modifies (\ref{selfcon}) to
\begin{equation}
Q ( i \omega_n ) = g \left(\omega_n^2 + \lambda - g J^2 Q ( i \omega_n )
+ \Sigma(i\omega_n )/ M \right)^{-1}.
\label{sigma}
\end{equation}
The function $\Sigma(\tau)$ is itself a non-linear functional of $Q(\tau)$,
obtainable by a $1/N$ expansion of $Z_0$.
Let us consider first the critical point $g=g_c$ and use the $M=\infty$
result $Q^{aa} ( i\omega_n ) \sim
|\omega_n |$ at low frequencies. The leading term in $\Sigma$ satisfies
$\mbox{Im} (\Sigma ( \omega + i 0^{+} )) \sim \omega^5$,
$\mbox{Re} (\Sigma ( \omega + i 0^{+} )) \sim a_1 + a_2 \omega^2$, at small
$\omega$; the suppression at low-frequencies in $\mbox{Im}(\Sigma)$
arises from restriction
in the phase space to three spin-wave decay. On the imaginary frequency
axis, this implies that the leading {\em non-analytic\/} term in
$\Sigma ( i \omega_n )$ is $\sim |\omega_n|^{5}$.
Now consider the self-consistency (\ref{sigma}). The analytic terms in
$\Sigma$ lead to apparently
innocuous frequency and mass renormalizations, while the
non-analytic terms vanish so rapidly that they don't modify the
assumed low-frequency
form $Q^{aa} ( i\omega_n ) \sim
|\omega_n |$; our initial assumption is therefore self-consistent.
Terms higher-order in $1/M$ have even weaker non-analytic contributions.
Thus there are
no modifications
to the critical properties, order-by-order to all orders in $1/M$.
Similar considerations also apply to the low-frequency spectrum
in the spin-fluid phase, where again the $M=\infty$ form survives.

This behavior can be better understood using a classical statistical mechanics
point of view, in which the system is viewed as a classical one dimensional
spin system with a long range interaction
$Q^{aa}(\tau)$; having solved the
model we can then require the self-consistency (\ref{qab}).
Our results above imply that the critical point of the quantum phase transition
corresponds to a spin system with $Q (\tau ) \sim 1/\tau^2$ for large $\tau$
($1/\tau^2$ is the Fourier-transform of $|\omega|$).
We may consider a lattice
discretization of $\tau$, and
also replace the fixed length spins by $M$-component soft spins $\vec{S}$ with
a
Landau-Ginzburg potential local in time. Thus we are led to a model with action
whose continuum limit is
\begin{eqnarray}
S&=&-\int d\tau\,d\tau'\,Q(\tau-\tau')\vec{S}(\tau)\cdot\vec{S}(\tau')
\nonumber\\
& &~~~~~~+\frac{1}{2}\int d\tau\,\left[
\frac{1}{g}(\partial_\tau\vec{S}(\tau))^2+
r\vec{S}^2(\tau)+u(\vec{S}^2)^2\right].
\label{LGaction}
\end{eqnarray}
where $g,r,u$ are constants.
This classical spin system, with $Q(\tau) \sim 1/\tau^{1+\sigma}$,
was studied many years ago~\cite{dyson}.
These authors found a high temperature paramagnetic phase with power-law
spin correlations, and a transition to a low-temperature ordered state
if $\sigma < 1$, or if $M=1$, $\sigma=1$. In the high temperature phase
they found $\left\langle \vec{S} (\tau) \cdot\vec{S}(\tau')\right\rangle
\sim 1/\tau^{1+\sigma}$ which is also the result obtained from the leading
term in the high-temperature expansion (expansion in powers of $Q$).
Throughout the high-temperature phase the
spin-spin correlation exponent is unmodified
by higher order terms.
As $Q(\tau)$ and $\left\langle \vec{S} (\tau) \cdot\vec{S}(\tau')\right\rangle$
have the same asymptotic decay, it is evident that
the self-consistency (\ref{qab}) can be satisfied for any value of
$\sigma$.
The result that
$\sigma=1$ corresponds to the quantum phase transition,
can be traced to the $\omega_n^2$ in
(\ref{selfcon}) or (\ref{LGaction}) which is generically present as the
leading analytic $\omega_n$
dependence. It thus has nothing to do with the critical point of the
one dimensional system;
the quantum-critical point corresponds to a point in the high-temperature
phase of the classical spin model.
With the choice $\sigma=1$, the other critical properties then follow; the
logarithmic violation of scaling comes in this model from summing tadpole
diagrams in the $S^4$ interaction. These arguments are valid
for all $M$ including $M=1$ (the transverse Ising case).

This paper has presented a soluble model with infinite-range interactions
which displays
a quantum phase transition
from a spin-glass to a spin fluid phase. Further theoretical work on the
extension of these results to finite-range interactions is clearly required.
Our result for $\chi^{\prime\prime} ( \omega ) $ in (\ref{chires}) has
qualitative similarities to the experimental results in the
transverse-field Ising magnet $Li Ho_x Y_{1-x} F_4$~\cite{rosen},
and the outlook for an eventual complete understanding of this system is
bright.

We thank A. Georges, D. Huse, J. Miller, and A.P. Young for useful discussions.
This research was supported by NSF Grants Nos. DMR 8857228, DMR 9157484
and by the A.P. Sloan Foundation.

\figure{
Phase diagram of $H$ in the $T$-$g$ plane at $M=\infty$. The line $g=0$
corresponds to the classical model of Ref~\cite{almeida}.
The quantum-disordered phase is the paramagnet at $T=0$. Regions in which
the spin fluctuations are primarily thermal/quantum are noted.
\label{phase}}
\end{document}